\documentclass[12pt]{book}
\input epsf
\input rotate
\makeatletter
\def\nothing#1{}
\newdimen\earraycolsep
\def\eqnarray{\stepcounter{equation}\let\@currentlabel\theequation
\global\@eqnswtrue\m@th
\global\@eqcnt\z@\tabskip\@centering\let \\\@eqncr
$$\halign to\displaywidth\bgroup\@eqnsel\hskip\@centering
$\displaystyle\tabskip\z@{##}$&\global\@eqcnt\@ne
\hskip 2\earraycolsep \hfil$\displaystyle{##}$\hfil
&\global\@eqcnt\tw@ \hskip 2\earraycolsep $\displaystyle\tabskip\z@{##}$\hfil
\tabskip\@centering&\llap{##}\tabskip\z@\cr}
\renewcommand{\theequation}{\arabic{equation}}
\renewcommand{\thetable}{\arabic{table}}
\renewcommand{\thefigure}{\arabic{figure}}

\def\title{\chapter}
\renewcommand\chapter{\ifodd\c@page\clearpage\else\cleardoublepage\fi
		    \global\@topnum\z@
		    \@afterindenttrue
		    \secdef\@chapter\@schapter}
\def\@makechapterhead#1{%
  \vspace*{120\p@}%
  {\parindent \z@ \raggedright \reset@font
    \bfseries #1\par
    \nobreak
    \vskip 36\p@
  }}
\def\author#1{{\pretolerance=10000 \raggedright \advance \leftskip by 1in \noindent #1 \vskip 1pc}}
\def\affiliation#1{{\advance\leftskip by 1in \noindent #1 \vskip -1pc}}

\renewcommand\section{\@startsection{section}{1}{\z@}{2pc \@plus 1ex minus
    .2ex}{1pc \@plus .2ex}{\reset@font\normalsize\bfseries}}
\renewcommand\subsection{\@startsection{subsection}{2}{\z@}{1pc \@plus 1ex
    minus.2ex}{1pc \@plus .2ex}{\reset@font\normalsize\bfseries}}
\renewcommand\subsubsection{\@startsection{subsubsection}{3}{\parindent}
	{1pc \@plus 1ex minus.2ex}{-0.5em}{\reset@font\normalsize\bfseries}}

\def\AmS{{\protect\the\textfont2 A\kern-.1667em\lower.5ex\hbox{M}\kern-.125emS}}

\def\p@LaTeX{{\family{times}\series{m}\shape{n}\selectfont L\kern-.36em\raise.3ex\hbox{\scriptsize A}\kern-.15em
T\kern-.1667em\lower.7ex\hbox{E}\kern-.125emX}}

\newlength{\colwidth}

\def\@oddhead{\hfil}
\def\@evenhead{\hfil}
\def\@oddfoot{{\bfseries\hfil\thepage}}
\def\@evenfoot{{\bfseries\thepage\hfil}}
\def\fnum@figure{\footnotesize\raggedright{\bfseries \figurename~\thefigure.}}
\def\fnum@table{\normalsize\raggedright{\bfseries \tablename~\thetable.}}
\long\def\@makecaption#1#2{\vskip 10\p@ {#1 #2\par}}
\long\def\@makefntext#1{\setbox0=\hbox{$\m@th^{\@thefnmark}$}\noindent\hangindent=\wd0 \box0 #1}
\def\centerfig#1#2#3#4{\vspace*{#2}\relax\centerline{\hbox to#1{\special{#4:#3.#4 x=#1, y=#2}\hfil}}}
\newbox\@atbox
\long\def\atable#1#2#3{\begin{table}[tbp]\centering\footnotesize
\setbox\@atbox\hbox{#2}
\parbox{\wd\@atbox}{\caption{#1}}\par\smallskip
#2
\par\smallskip\parbox{\wd\@atbox}{\raggedright #3}
\end{table}}
\def\@bibitem{\par\noindent \hangindent=2pc \hangafter=1}
\def\thebibliography{%
\section*{REFERENCES}%
\bgroup\footnotesize
\def\newblock{\hskip .11em plus.33em minus.07em}%
\let\bibitem\@bibitem}
\def\endthebibliography{\par\egroup}
\def\@nbibitem#1{\par\noindent \hangindent=2pc \hangafter=1
\refstepcounter{enumi}\hbox to 2pc{\arabic{enumi}.\hfil}%
\immediate\write\@auxout{\string\bibcite{#1}{\arabic{enumi}}}}
\def\numbibliography{%
\section*{REFERENCES}%
\bgroup\footnotesize
\setcounter{enumi}{0}%
\def\newblock{\hskip .11em plus.33em minus.07em}%
\let\bibitem\@nbibitem}
\def\endnumbibliography{\par\egroup}
\def\@cite#1#2{{#1\if@tempswa , #2\fi}}
\makeatother

\begin{document}
\chapter
{Ensemble density functional theory for inhomogeneous fractional quantum
Hall systems}
\author{O. Heinonen, M.I. Lubin, and M.D. Johnson}
\affiliation{Department of Physics, University of Central Florida,
\\Orlando, FL 32816-2385
}

\section{Introduction}
The fractional quantum Hall effect (FQHE) occurs in a two-dimensional
electron gas (2DEG) in a strong magnetic 
field oriented perpendicular to the plane of the 
electrons[\cite{Girvin}]. The effect was discovered as a transport anomaly. 
In a transport
measurement 
it is noted that at certain strengths $B^*(n)$, which depend on the
density $n$ of the 2DEG, current can flow without any dissipation. 
That is, there
is no voltage drop along the flow of the current. At the same time, the
Hall voltage perpendicular to both the direction of the current and of the
magnetic field is observed to attain a quantized value for a small, but finite,
range of magnetic field or density, depending on which quantity is varied
in the experiment. 
The effect is understood to be the result of an excitation gap in the
spectrum of an infinite 2DEG at these magnetic fields. A convenient 
measure of the density of a 2DEG in a strong magnetic field is given by
the filling factor $\nu=2\pi\ell_B^2 n$, with 
$\ell_B=\sqrt{\hbar c/(eB)}$ the magnetic length. The filling factor 
gives the ratio of the number of particles to the number of available states
in a magnetc sub-band (Landau level), or, equivalently, the number of particles per
flux quantum $\Phi_0=hc/e$.
The quantum Hall effect was first discovered[\cite{Klitzing}]
at integer filling factors. In this
integer quantum Hall effect, the energy gap is nothing but the kinetic
energy gap $\hbar\omega_c=\hbar eB/(m^*c)$.
Later, the fractional quantum Hall effect (FQHE)
was discovered[\cite{Stormer}] 
at certain
rational filling factors of the form $\nu=p/q$, with $p$ and $q$ relative
primes, and $q$ odd. In the FQHE, the excitation gap is a consequence of
the strong electron-electron correlations. Therefore, any computational
approach to the quantum Hall effect must accurately treat the 
electron correlations in order to capture the FQHE {\em at all}. 
We emphasize that this fact makes any 
density functional approach to the FQHE 
qualitatively different from `standard' applications to atomic, 
molecular or condensed matter systems. In such standard 
applications, the electron
correlations usually give a quantitatively important correction of
the order of perhaps 10 -- 20\% but do not usually {\em qualitatively}
change the results in a fundamental way. Furthermore, standard applications
of density functional theory do not require ensemble density functional
theory (in such applications, the systems under considerations are
typically pure-state $v$-representable). Our work represents
(to the best of our knowledge) the first practical applications of
ensemble density functional theory to a system which is {\em not}
pure-state $v$-representable.

\begin{figure}[t]
\setbox0=\hbox{
\epsfxsize 13.0 cm
\epsfbox{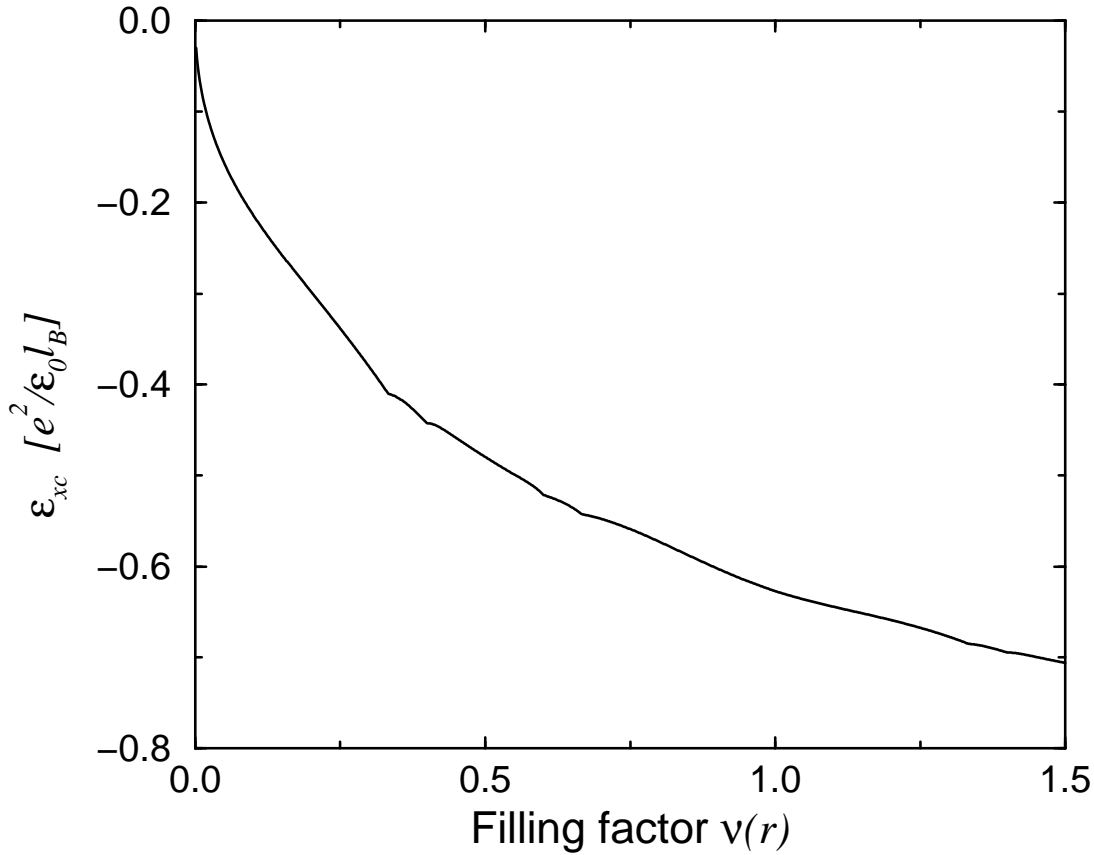}
}
\rotr0
\caption{ The ground state energy $\epsilon_{\rm xc}$ per particle
of an infinite, homogeneous, spin-polarized FQHE system is depicted 
as a function of
filling factor. The cusps at $\nu=1/3,2/5,3/5,2/3,4/3$, and $\nu=7/5$ are included. 
In this
plot, the smooth part of the exchange-correlation energy was obtained from
the Levesque-Weiss-MacDonald interpolation formula Eq. (\ref{eq:exc1})}
\label{energy:fig}
\end{figure}

Our understanding of the origin of the FQHE started with Laughlin's
seminal paper of 1983[\cite{Laughlin}], which dealt with the simplest
fractions $\nu=1/m$, with $m$ an odd integer. At these values of $\nu$, there
are on the average $m$ magnetic flux quanta $\Phi_0=hc/e$ per electron. 
Laughlin constructed a variational
wavefunction for spin-polarized systems in strong magnetic fields, strong enough
that
the splitting $\hbar\omega_c$ between the magnetic subbands, 
or Landau levels, can be taken to be infinite. The wavefunction can then
be constructed from single-particle states entirely within the lowest
Landau level. Laughlin wrote the variational wavefunction as
\begin{equation}
\Psi_m\propto\prod_{i,j}
\left(z_i-z_j\right)^m\exp\left[-\frac{1}{4}\sum_k|z_k|^2
\right]
\label{Laughlin:eq}
\end{equation}
where $z_j=x_j+iy_j$ is the coordinate of the $j$th
electron in complex notation. This wavefunction is an eigenstate of
angular momentum. Laughlin demonstrated that the
system having the wavefunction Eq. (\ref{Laughlin:eq}) is an incompressible
liquid with $\nu=1/m$, $m$ odd,  
with an energy gap to excitations, and that the elementary excitations are
fractionally charged quasi-holes or quasi-particles of charge $e^*=\pm e/m$.
The origin of the energy gap can be understood in the so-called
pseudo-potential representation of the electron-electron interactions
[\cite{Haldane_pseudo}].
Here, the electron-electron interaction $V({\bf r}_i-{\bf r}_j)$ 
between electrons $i$ and $j$ is
decomposed into strengths $V_\ell$ in relative angular momentum channels 
$\ell=0,1,2,\ldots$,
of the two electrons. For any realistic interaction $V({\bf r}_i-{\bf r}_j)$,
it turns out that $V_0>V_1>V_2>\ldots$.
Consider the case $\nu=1/3$. In this case, 
the lowest angular momentum pseudo-potential that enters
into the $m=3$ Laughlin description
is $V_1$, the interaction energy of two electrons of unit 
relative angular momentum (in units of $\hbar$).
(Only odd relative angular momenta are permissible for 
spin-polarized electron wavefunctions,
since they have to be anti-symmetric under 
interchange of electron coordinates.)
The Laughlin wavefunction is a very cleverly
constructed highly correlated state which
completely excludes unit relative angular momentum between
any two electrons, and is furthermore the only state which satisfies this
property at $\nu=1/3$.  Therefore, any excited state at this filling factor
must contain some
electrons with unit relative angular momentum. The energy gap is due to
the cost of this, and hence is of order $V_1$. 
Figure \ref{energy:fig} depicts the exchange-correlation
energy per particle for infinite, homogeneous FQHE systems {\em vs.}
filling factor. The cusps at 
filling factors $\nu=1/3,2/5,3/5,2/3,4/3$, 
and $\nu=7/5$ have been included to scale.
Note that these cusps are barely visible on this scale, yet they are
responsible for all the physics of the FQHE!

\section{Finite systems, and spins}
We have outlined above how the
electron-electron interactions in an infinite, homogeneous system
produce the excitation gap. It is important to note that these gaps are only
for excitations in the {\em bulk} of the system. 
When a system is bounded there {\em must} be gapless excitations located
at the boundaries of the system[\cite{Halperin,MacDonald_arg}]. Since all
experimental systems are finite and inhomogeneous, the 
low-energy properties probed by experiments are necessarily
determined by the gapless edge excitations. Advances in 
semiconductor nanofabrication technologies have lead to the possibility
of manufacturing systems which are extremely inhomogeneous, and in practice
dominated by edges. As an example, recent experiments
have even been performed on tiny quantum dots, containing about 30 
electrons[\cite{McEuen,Klein}]. Also, edge structures in inhomogeneous
systems have been studied directly using capacitance
spectroscopy[\cite{Zhitenev1}], time-resolved measurements of
edge magnetoplasmons (the gapless modes along the boundary)[\cite{Zhitenev2}], 
and by surface acoustic wave techniques
capable of resolving very small inhomogeneities in the electron
density[\cite{Shilton}]. 

So far, we have limited the discussion to spin-polarized system. At first look
it seems reasonable that in the strong magnetic fields used in
experiments on QHE systems the Zeeman splitting $g\mu_B B$, where
$g$ is the effective Land\'e factor and $\mu_B$ the Bohr magneton, is
large enough that the high-energy spin direction is energetically
inaccessible. However, two factors conspire to make the Zeeman
splitting very low in GaAs systems. First of all, spin-orbit coupling
in the GaAs conduction band effectively lowers the Land\'e factor to
$g\approx0.44$. Second, the low effective mass, $m^*\approx0.067m_e$, 
further reduces the ratio of Zeeman energy to cyclotron energy to
about 0.02, compared to its value of unity for free electrons. 
For magnetic fields of about 1 -- 10 T, the Coulomb energy scale of the
electron-electron interactions, $e^2/(\epsilon_0\ell_B)$, where
$\epsilon_0\approx12.4$ is the static dielectric constant, is of the
same order as the cyclotron energy. As a first approximation, one should
then set the Zeeman energy to zero, rather than infinite, since it is
two orders of magnitude smaller than the other energy scales. As a consequence,
the spin-degree of freedom is governed by the electron-electron
interactions, rather than by the Zeeman energy. This dramatically
changes the nature of the low-energy bulk single-particle excitations
near filling factors of the form
$\nu=1/m$, with $m$ odd,
from single-particle spin-flips to charge-spin textures.
In these objects, loosely called skyrmions, 
the spin density varies smoothly over a distance of
several magnetic lengths, so that the system can locally take advantage of the
exchange energy.
Skyrmions in the quantum Hall effect were first predicted by Sondhi 
{\em et al.\/}[\cite{Sondhi}]. Recent experiments[\cite{Barrett}] 
measuring the
spin-polarization near $\nu=1$ have provided evidence for this. According
to calculations by Karlhere {\em et al.\/}[\cite{Kivelson}], the skyrmion
excitations in an infinite homogeneous system may lead to charge-spin
textured appearing at the edges of QHE systems, which would
dramatically change our understanding of QHE edges.

In order to accurately understand the experiments and inhomogeneous
FQHE systems in general, we must have a way of accurately calculating their
properties, including the spin degrees of freedom.
Certain aspects of inhomogeneous FQHE systems have been studied by different
techniques. For example, field theories can be constructed to study the
low-energy limit of the gapless edge excitations for spin-polarized
systems[\cite{Wen}], or the spin texture on the edge of an infinite
Hall bar[\cite{Kivelson}]. However, effective field theories
typically do not give accurate quantitative results, other than for {\em e.g.}
critical exponents near fixed points. So-called composite
fermion methods have been used for non-interacting composite 
fermions[\cite{Jain_Kawamura}] and in a Hartree approximation to study
finite spin-polarized FQHE systems[\cite{Brey,Chklovskii2,Renn}]. 
In this approach, the Chern-Simons term, arising from
the singular gauge transformation, is replaced by its smooth spatial
average, and the composite fermion mass, which is not well known, has to
be put in by hand. The electron-electron interaction
can then treated in a self-consistent
Hartree approximation. The hope is then that the most important aspects of the
electron-electron correlations are included in this approximation.
Near $\nu=1$, at which the Slater determinant $\Psi_1$ 
is the exact ground state, it makes sense to use the Hartree-Fock approximation,
and the stability of a spin-polarized 
quantum dot at $\nu=1$ as a function of confining
potential has been studied in this approximation[\cite{MacD_Yang,Chamon}]. 
The edge structure of spin-polarized systems 
has also been studied using semiclassical 
methods[\cite{Beenakker,Chklovskii}], in 
which the electron-electron interaction is included at the Hartree level and
it is furthermore assumed that all potentials vary on a length scale much
larger than $\ell_B$. Beenakker[\cite{Beenakker}], 
and Chklovskii, Shklovskii and Glazman[\cite{Chklovskii}] 
demonstrated that the edge of an
integer quantum Hall system, in which the correlation energies between the
electrons can be ignored, consists of a sequence of compressible and
incompressible strips. The widths of the incompressible strips
is determined by the length over which the effective confining potential
(external plus Hartree) varies an amount equal to the energy gap
$\hbar\omega_c$. The origin of the compressible and incompressible
strips are the energy gaps, which are the kinetic energy gaps $\hbar\omega_c$
in the case of the integer quantum Hall effect. But it is easy to generalize
the argument to include the energy gaps causing the FQHE[\cite{Chklovskii}].
The conclusion is then that there should be compressible and incompressible
strips, with the density of each incompressible strip fixed at the value
of an FQHE fraction.
The width of each incompressible
strip is then fixed by the length over which the effective confining potential
varies an amount equal to the energy gap of the FQHE fraction corresponding to
the density of that strip.

Quantum dots including the spin degree of freedom 
have also been studied by direct numerical
diagonalizations[\cite{Yang_MacDonald_Johnson}]. These calculations
demonstrated the importance of the spin degree of freedom.
At the present, numerical diagonalizations are limited to systems with
of the order of 10 electrons. 

It is highly desirable to have a 
computational approach which accurately includes electron-electron
correlations and spin degree of freedoms, and which 
can handle inhomogeneous systems  
with on the order of 
$10^2$--$10^3$ electrons.
One such approach which is in principle valid
for any interacting electron system is 
density functional theory (DFT)[\cite{Kohn_Vashista,Dreizler,Parr}]. 
There have been some attempts to apply density functional theory to the
FQHE. Ferconi and Vignale[\cite{Ferconi_CDFT}] applied current density
functional theory[\cite{Vignale_CDFT}] to small, parabolically confined
quantum Hall systems and showed that the current density functional theory gave
good results for the ground state energy and spin polarization near $\nu=1$.
However, the energy gaps due to correlation effects were not included
in that calculation. Ferconi, Geller and Vignale[\cite{Ferconi}] 
also recently studied spin-polarized
FQHE systems within the spirit of the DFT using an extended 
Thomas-Fermi approximation at low, but non-zero, temperatures. 
In this, the kinetic energy was treated as
a local functional, as in the standard Thomas-Fermi approximation, while the
exchange-correlation energy was included 
in a local density approximation (LDA). 
This extended Thomas-Fermi approximation is valid in the limit
of very slowly varying confining potential. Ferconi, Geller and Vignale
focused on the incompressible and compressible strips at an edge of an
FQHE system, and obtained results in agreement with the predictions
by Chklovskii, Shklovskii, and Glazman[\cite{Chklovskii}].

We have developed for the fractional quantum Hall effect 
an ensemble DFT scheme within the local density 
approximation, 
and have applied it to 
circularly
symmetric quantum dots[\cite{Heinonen1,Heinonen2}]. 
In our approach, the kinetic
energy is treated exactly, and the density represented by Kohn-Sham 
orbitals.
The results are in good agreement with results
obtained by semiclassical[\cite{Beenakker,Chklovskii,Ferconi}], 
Hartree-Fock[\cite{MacD_Yang,Chamon}] (for cases where the
correlations do not play a major role), and exact diagonalization 
methods[\cite{Yang_MacDonald_Johnson}].
Our calculations for spin-polarized systems show that the exchange and
correlation effects of the FQHE are very well represented by the LDA and that
our approach provides a computational scheme to model large inhomogeneous
FQHE systems. We note that
there exist previous formal DFTs 
for strongly correlated systems, in particular for high-temperature
superconductors [\cite{Gross}], and DFT calculations of high-T$_{\rm c}$
materials [\cite{Pickett}] and transition-metal oxides [\cite{Svane}].
However, ours are, to the
best of our knowledge, the first practical LDA-DFT calculations of
a strongly correlated system in strong magnetic fields, and demonstrate the
usefulness of the LDA-DFT in studying large inhomogeneous FQHE systems.
Recently, we have generalized our DFT approach to include spin
degrees of freedom in an approximation in which the spin-quantization
axis is parallel to the external magnetic field. While such an
approximation cannot capture skyrmion-like excitations, in which the
spin quantization axis is tumbling in space, it does demonstrate the
existence of spin structures near edges of inhomogeneous systems
consistent with numerical
diagonalizations[\cite{Yang_MacDonald_Johnson}] and
Hartree-Fock calculations[\cite{Kivelson}].
We are presently working on extending the DFT approach to allow for
the possibility of a tumbling spin quantization axis.

\section{Ensemble density functional theory approach for spin-polarized systems}
In typical DFT calculations of systems of $N_{\rm el}$ electrons, 
the standard Kohn-Sham (KS) scheme[\cite{KohnSham}] is implemented, 
in which the particle density $n({\bf r})$ is expressed in terms of
a Slater determinant of $N\geq N_{\rm el}$ 
KS orbitals, $\psi_{\alpha}({\bf r})$. These obey an effective
single-particle Schr\"odinger equation 
$H_{\rm eff}\psi_\alpha=\epsilon_\alpha\psi_\alpha$, which is solved 
self-consistently by occupying the $N_{\rm el}$ KS orbitals with the lowest
eigenvalues $\epsilon_\alpha$,
and iterating. 
This scheme works well in practice
for pure-state $v$-representable systems, 
for which the true 
electron density can be represented by a single Slater
determinant of single-particle wavefunctions. 
However, when the KS orbitals are degenerate at the Fermi energy
(which we identify with the largest $\epsilon_\alpha$ of the occupied
orbitals)
there is an
ambiguity in how to occupy these degenerate orbitals.
There exists an extension of DFT which is formally able to deal
with this situation. This extension is called ensemble DFT[\cite{Dreizler,Parr}], 
and in it,
the density of the system is represented by an ensemble of Slater
determinants of KS orbitals. However, while it can be  shown using 
ensemble DFT 
that such a representation of the density is rigorous, it cannot be shown
{\em how} the degenerate KS orbitals at the Fermi energy should be occupied,
{\em i.e.,\/} there has not been available a practical computational scheme
for ensemble density functional theory.  

We begin by first demonstrating that fractional quantum Hall systems
are not in general pure-state $v$-representable, which means their densities
cannot be obtained by a single Slater determinant of Kohn-Sham orbitals.
Hence, an ensemble representation has to be used. Consider a 
uniform fractional quantum Hall system at filling factor $\nu=1/3$.
This density can be obtained by forming a density matrix of the form
\begin{equation}
\hat D=\frac{1}{3}\sum_{i=1}^3 |\Psi_i\rangle\langle\Psi_i|.
\label{eq:density_matrix}
\end{equation}
Here, $|\Psi_i\rangle$ are the three possible degenerate Slater determinants
obtained by filling every third single-particle orbital in the
lowest Landau level. Using momentum eigenfunctions,
we can write in occupation-number representation
$|\Psi_1\rangle=\{100100100\ldots\}$, $|\Psi_2\rangle=\{010010010\ldots\}$,
and $|\Psi_3\rangle=\{001001001\ldots\}$. The corresponding ensemble
density is then
\begin{equation}
n_D({\bf r})={\rm Tr}\left\{
\hat D\hat n({\bf r})
\right\}
=\frac{1}{3}\sum_{i=1}^3n_i({\bf r})={1\over3\times2\pi\ell_B^2},
\label{eq:ens_density}
\end{equation}
that is, the filling factor is fixed at $\nu=1/3$. Because we can construct
the correct ground state density from a density matrix of the form
Eq. (\ref{eq:density_matrix}) with $q>2$, it follows from a theorem
by Levy[\cite{Levy}] and Lieb[\cite{Lieb}] that this system then has a density
which cannot be derived from a single ground state Kohn-Sham
determinant, {\em i.e.,}, $n_D({\bf r})$ is not pure-state 
$v$-representable. (This is true whenever we can write the ground
state density of a system in the form of Eq. (\ref{eq:density_matrix}) with more
than two terms.) However, $n_D({\bf r})$ is still associated with
an external potential, and is ensemble representable.

Although ensemble DFT has been developed formally, there are in practice
few examples of applications and calculations using ensemble DFT
for ground state calculations. A significant aspect of our
work is that we have developed an ensemble scheme which is practical 
and useful
for the study of the FQHE.
In ensemble DFT, any physical density
$n({\bf r})$ can be represented by
$
n({\bf r})=\sum_{mn}f_{mn}|\psi_{mn}({\bf r})|^2,
$
where $f_{mn}$ are occupation numbers satisfying
$0\leq f_{mn}\leq1$, and the orbitals $\psi_{mn}$
satisfy the equation
\begin{equation}
\left\{
\frac{1}{2m^*}\left[{\bf p}+\frac{e}{c}{\bf A}({\bf r})\right]^2
+V_{\rm ext}({\bf r})+V_{\rm H}({\bf r})
+V_{\rm xc}({\bf r},{\bf B})\right\}\psi_{mn}({\bf r})=\epsilon_{mn}
\psi_{mn}({\bf r}),\label{HK}
\end{equation}
where $\nabla\times{\bf A}({\bf r})={\bf B}({\bf r})$.
In equation (\ref{HK}),  
$V_{\rm H}({\bf r})$ is the 
Hartree interaction of the 2D electrons, and, as usual, 
$V_{\rm xc}({\bf r},{\bf B})$ is the exchange-correlation potential, defined as
a functional derivative of the exchange-correlation energy 
$E_{\rm xc}[n({\bf r}),{\bf B}]$ 
of the system with respect to density:
$
V_{\rm xc}({\bf r},{\bf B})={\delta E_{\rm xc}[n({\bf r}),{\bf B}]
\over\delta n({\bf r})}.
$
(We will hereafter not explicitly indicate the parametric dependence
of $V_{\rm xc}$ and
$E_{\rm xc}$ on $\bf B$.) For the case of the FQHE, we know that the
exchange-correlation potential will be crucial, as it contains all the effects
of the electron correlations which cause the FQHE in the first place, and
a major part of the DFT application is to come up with an accurate model
of $E_{\rm xc}$ and so of $V_{\rm xc}$. Leaving this question aside for a moment,
and assuming that we have succeeded in doing so,  
the practical question is then how to determine
the KS orbitals and their occupancies in the presence of degeneracies. We
have devised an empirical scheme, which produces a set of
occupancies for the KS orbitals which satisfy some minimum requirements, namely
(a) the scheme converges
to physical densities (to the best of our knowledge) for FQHE systems, 
(b) it reproduces finite temperature DFT distributions at finite
temperatures, and (c) it reproduces the standard Kohn-Sham scheme for
systems whose densities can be represented by a single Slater determinant.

In our
scheme, we start with input occupancies and single-particle orbitals and
iterate the system $N_{\rm eq}$ times using the KS scheme. 
The number $N_{\rm eq}$ is chosen large enough (about 20--40 
in practical calculations)
that the density is close to the final density after $N_{\rm eq}$ iterations. 
If the density of the system could be represented by a single
Slater determinant of the KS orbitals, we would now essentially
be done. However, in this system there are now in general many degenerate
or near-degenerate orbitals at the Fermi energy. After each iteration, the
Kohn-Sham scheme chooses to occupy the $N_{\rm el}$ orbitals with the lowest
eigenvalues, corresponding to making a distinct Slater determinant of these
orbitals. But there will be small fluctuations in the density between each
iteration, which cause a different subset of these (near) degenerate
orbitals to be occupied after each iteration.
This corresponds to constructing
different Slater determinants after each iteration, and the occupation numbers
$f_{mn}$ of these orbital are zero or unity more or less at random after
each iteration. This means that the computations will never converge.
However, the {\em average} occupancies, {\em i.e.,\/} the
occupancies averaged over many iterations, become well defined and approach
a definite value, {\em e.g.,\/} 1/3 for orbitals localized 
in a region where the local filling factor is close to $\nu=1/3$.
Therefore, we use these average occupancies to construct an ensemble by
accumulating running average occupancies $\langle f_{mn}\rangle$
after the initial $N_{\rm eq}$ iterations 
\begin{equation}
\langle f_{mn}\rangle=
{1\over(N_{\rm it}-N_{\rm eq})}\sum_{i=N_{\rm eq}+1}^{N_{\rm it}}f_{mn,i},
\label{average_occ:eq}
\end{equation}
where $f_{mn,i}$ is the occupation number (0 or 1) of orbital $\psi_{mn}$ 
after the $i$th iteration,
and use these to calculate densities. 
Thus, our algorithm essentially
picks a different (near) degenerate Slater determinant after each 
iteration, and these
determinants are all weighted equally in the ensemble.
It is clear that this scheme reduces to the KS scheme for which the
density can be represented by a single Slater determinant of KS orbitals
(for which the KS scheme picks only the one Slater determinant which gives
the ground state density)
for $N_{\rm eq}$ large enough. Moreover, we have numerically 
verified that a finite-temperature version of our scheme converges
to a thermal ensemble at finite temperatures down to
temperatures of the order of $10^{-3}\hbar\omega_c/k_B$. We
have also performed some Monte Carlo simulations about the
ensemble obtained by our scheme. In these simulations, we used a Metropolis
algorithm to randomly change the occupation numbers about our converged
solution, keeping the chemical potential fixed. The free energy of the
new set of occupation numbers was calculated self-consistently. 
If the free energy decreased,
this set was kept, and if the free energy increased, the set was kept
if a random number was smaller than $\exp\left[-\Delta F/k_BT^*\right]$, where
$\Delta F$ is the change in free energy, and $T^*$ a fictitious temperature. 
The results were that to within numerical
accuracy our ensemble DFT scheme gives the lowest free energy.
As a condition for convergence, we typically demanded that the difference
between the input and output ensemble densities, $n_{\rm in}(r)$ and
$n_{\rm out}(r)$, of one iteration should satisfy
\begin{equation}
{1\over N_{\rm el}}\int_0^\infty\,\left|n_{\rm in}({\bf r})-n_{\rm out}({\bf r})
\right|
d{\bf r}<5\times10^{-4}.
\end{equation}

Practical density functional theory calculations hinge on the availability
of good approximations for the exchange-correlation potential $V_{\rm xc}$,
which enters in the effective Schr\"odinger equation for the KS orbitals.
The simplest, and probably the most commonly used, approximation is the
local density approximation (LDA).
In this approximation, the exchange-correlation
energy is assumed to be a {\em local function} of density, so that
the total exchange-correlation energy consists of contributions from the
local density of the system. Thus, in this approximation one writes
$
E_{\rm xc}/N=\int d{\bf r}\epsilon_{\rm xc}(\nu)n({\bf r}),
$
where $\epsilon_{\rm xc}(\nu)$ is the exchange-correlation energy per
particle in a {\em homogeneous} system of constant density 
$n=\nu/(2\pi\ell_B^2)$ and 
filling factor $\nu$. In other words, in the LDA one assumes that the
system is locally homogeneous, {\em i.e.,\/} the system can locally
be approximated to have the energy per particle of an infinite, homogeneous
system of the local density. This approximation obviously makes sense
if the density of the system varies on a very long length scale, while
it could be questionable for systems in which the density varies on
some microscopic length scale. However,  
experience has shown that the LDA often works surprisingly well,
even
for systems in which the electron density is strongly 
inhomogeneous[\cite{Kohn_Vashista}].
In the
FQHE, the length scale of exchange-correlation interactions 
and density fluctuations is
given by the magnetic length $\ell_B$ due to the Gaussian fall-off of
any single-particle basis in which the interacting Hamiltonian
is expanded. The densities are relatively smooth on this length scale,
which gives us additional hope that the LDA will work well for the FQHE, too.
In addition, the cusps in the exchange-correlation energy will suppress density
fluctuations, so in this sense one can actually expect the basic physics
of the FQHE to make the LDA a good approximation.

We construct our exchange-correlation energy by writing
\begin{equation}
\epsilon_{\rm xc}(\nu)=\epsilon_{\rm xc}^{\rm LWM}(\nu)+
\epsilon_{\rm xc}^{\rm C}(\nu).
\label{eq:exc1}
\end{equation}
Here, $\epsilon_{\rm xc}^{\rm LWM}(\nu)$ is a smooth interpolation formula
(due to Levesque, Weiss and Mac\-Donald[\cite{Levesque}])
between ground state 
energies at some rational fillings. The
second term, $\epsilon_{\rm xc}^{\rm C}(\nu)$, 
is all-important for the study of the FQHE. This term contains the
cusps in the ground state energy which cause the FQHE. Here we have used
a simple model which captures the essential physics. We model
$\epsilon_{\rm xc}^{\rm C}(\nu)$ by constructing it to be zero at
values of $\nu=p/q$ which display the FQHE. Near $\nu=p/q$, 
$\epsilon_{\rm xc}^{\rm C}(\nu)$ is linear and
has at $\nu=p/q$ a discontinuity in the
slope related to the chemical potential gap 
$\Delta \mu=q(|\Delta_p|+|\Delta_h|)$. Here $\Delta_{p,h}$ are the 
quasiparticle (hole) creation energies  which
can be obtained from the literature [\cite{Morf_Halperin,Morf_Ambrumenil}]
at fractions $\nu=p/q$.
Farther
away from $\nu=p/q$, $\epsilon_{\rm xc}^{\rm C}(\nu)$ decays to zero.
Finally, in
the LDA $V_{\rm xc}(r)$ is obtained from $\epsilon_{\rm xc}(\nu)$ as
$
V_{\rm xc}(r)=\left.{\partial \left[\nu\epsilon_{\rm xc}(\nu)\right]
\over\partial\nu}\right|_{\nu=\nu(r)}
$
at constant $B$. In our calculations, we restrict ourselves to include
only the cusps at $\nu=1/3,2/5,3/5$ and $\nu=2/3$ (and the analogous ones
at 4/3, 7/5, 8,5 and 5/3), which are the
strongest fractions. These are some of the fractions 
of the form $
\nu={p\over(2p\pm1)}$ generated by the  
so-called $V_1$-model, in which only the pseudo-potential $V_1$ is included.

\begin{figure}[t]
\setbox1=\hbox{
\epsfxsize 13.0 cm
\epsfbox{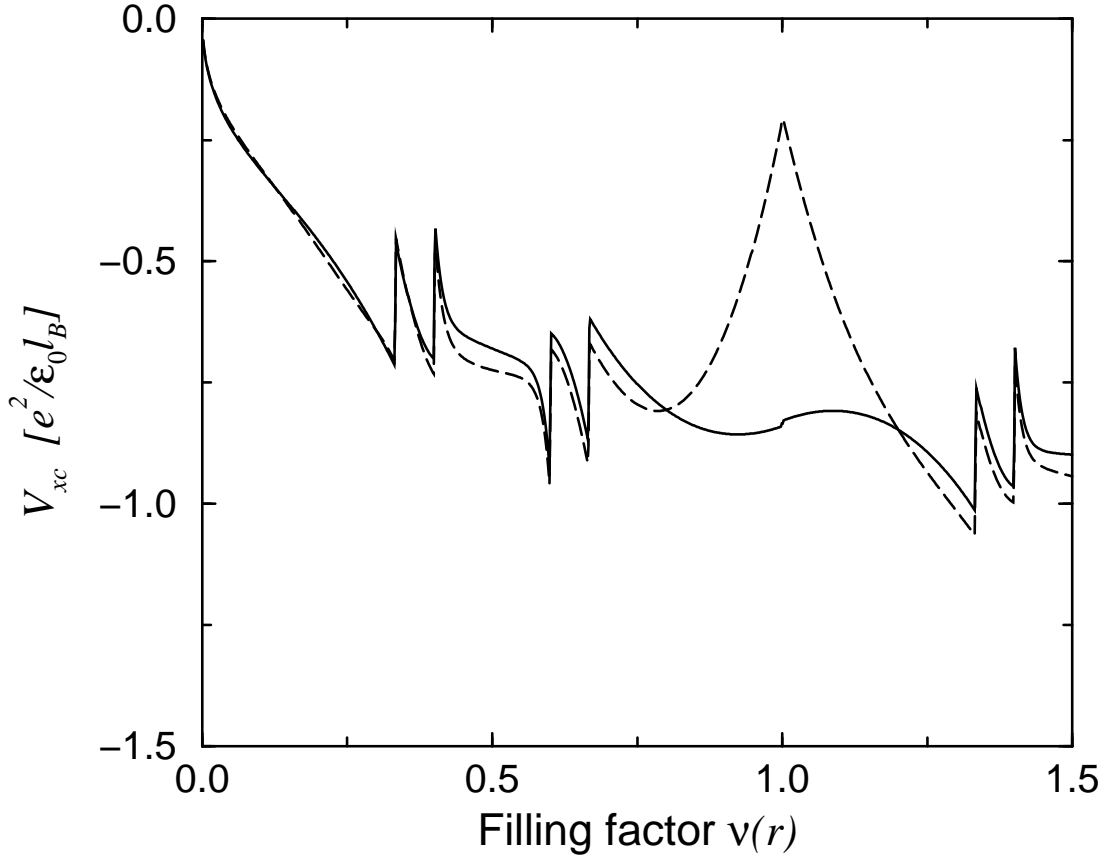}
}
\rotr1
\caption{Exchange-correlation potentials $V_{{\rm xc}\uparrow}$ (solid line)
and $V_{{\rm xc}\downarrow}$ (dashed line) for a spin-polarized system 
as function of
filling factor in units
of $e^2/(\epsilon_0\ell_B)$ for $0\leq\nu\leq 1.5$.
The increase in $V_{\rm xc}$
at an FQHE filling factoroccurs over a range of filling factor of 0.002.}
\label{V_xc}
\end{figure}

A technical difficulty arises in the LDA: the 
discontinuities in $V_{\rm xc}(r)$ in the LDA give rise to a 
numerical instability. The reason
is that an arbitrarily small fluctuation in charge density close to an FQHE
fraction gives rise to a finite change in energy. Imagine that
the local filling factor $\nu(r)$ 
in some neighborhood of a point $r$ is very close to,
but less than, say, 1/3 after one iteration. In this neighborhood, the 
local exchange-correlation potential will then form a potential well
with
sharp barriers at the points around $r$ where $\nu(r)=1/3$. During
the next iteration, charge will then be poured into this well. As a
result, the local filling factor will after this iteration exceed 
1/3, and in this neighborhood $V_{\rm xc}$ now forms a potential barrier of
finite height.
So in the next iteration, charge is removed from this
neighborhood, and so on. We can see that this leads to serious
convergence problems. To overcome this,
we made the compressibility of the system finite, but very small, 
corresponding to a finite, but very large, curvature instead of a point-like 
cusp in $\epsilon_{\rm xc}$ at the FQHE fractions. In other words,
instead of having a step-like discontinuity $\Delta\mu$ in the chemical potential,
it rises smoothly an amount $\Delta\mu$ over an interval $\gamma$ in the
filling factor. What we found worked very
well in practice was to have the discontinuity in chemical potential occur
over an interval of filling factor $\gamma$ of magnitude 
$10^{-3}$. This corresponds
to a sound velocity of about $10^6$ m/s in the electron gas, which is
three orders of magnitude larger than the Fermi velocity of a 2D electron
gas at densities typical for the FQHE. In general, the finite compressibility
does not lead to any spurious physical effects so long as the energy of
density fluctuations on a size of the order of the systems size is larger
than any other relevant energy in the problem. The only noticeable
effect is that incompressible plateaus, at which the density would be
perfectly constant were the compressibility zero, will have density
fluctuations on a scale of $\gamma$. 
Figure \ref{V_xc} depicts  $V_{\rm xc}$ for a spin-poalized system used in our calculations 
as a function of
filling factor.

\section{Applications to spin-polarized quantum dots}
We have self-consistently solved the KS equations Eqs. (\ref{HK})
for a spin-polarized quantum dot in a parabolic external potential, 
$V_{\rm ext}(r)=\frac{1}{2}m^*\Omega^2r^2$, by expanding the KS
orbitals $\psi_{mn}({\bf r})=e^{im\phi}\varphi_{mn}(r)$ in the eigenstates
of $H_0=\frac{1}{2m^*}\left({\bf p}+\frac{e}{c}{\bf A}({\bf r})\right)^2$.
We use the cylindrical gauge, ${\bf A}({\bf r})=\frac{1}{2}Br\hat\phi$, 
and include the four lowest Landau levels ($n=0,\ldots,3$). We
chose the static dielectric constant $\epsilon_0=12.4$, appropriate for
GaAs, and a confining potential of strength[\cite{McEuen}]
$\hbar\Omega=1.6$ meV.

\begin{figure}[t]
\setbox2=\hbox{
\epsfxsize 13.0 cm
\epsfbox{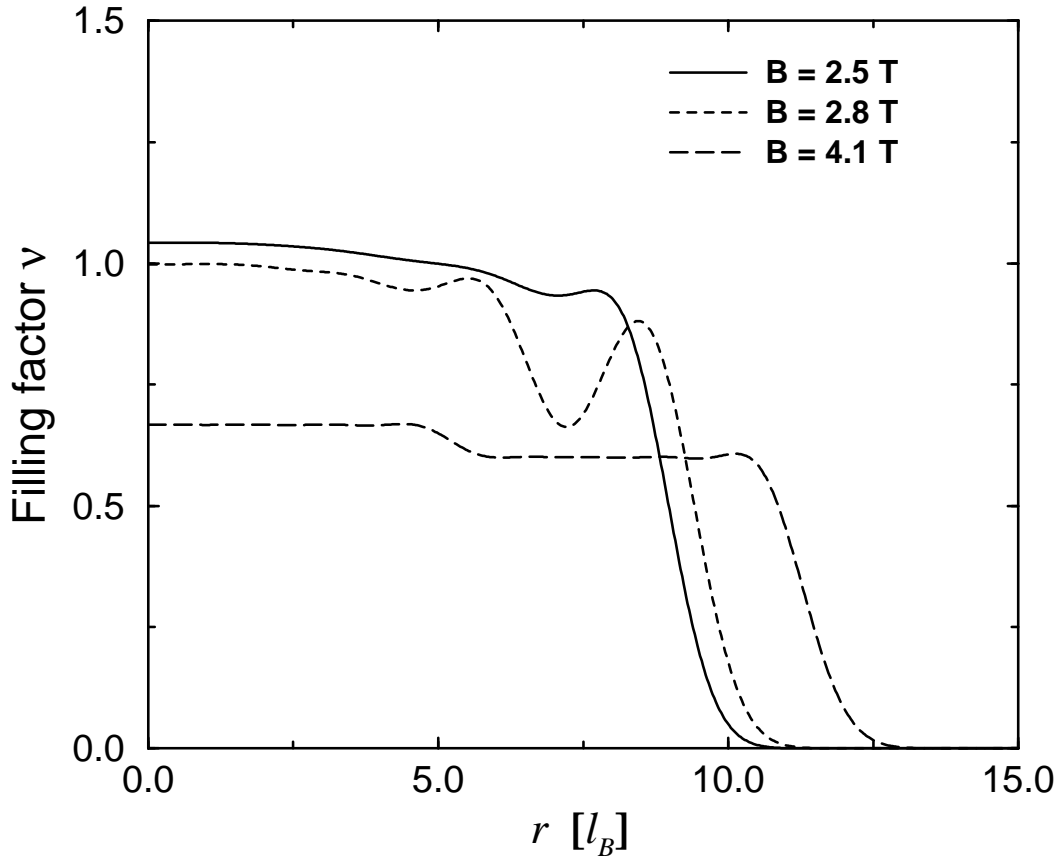}}
\rotr2
\caption{Edge reconstruction of a quantum dot as the magnetic field
strength is increased. 
Plotted here is the local filling factor 
$\nu(r)$ for a parabolic quantum dot with
$\hbar\Omega=1.6$ meV and
40 electrons. 
For magnetic field strengths $B< 2.5$ T the dot forms a maximum density
droplet, and for $B\approx2.8$ T, an exchange hole is formed. For stronger
magnetic fields, incompressible regions form, separated by compressible strips.}
\label{reconstr}
\end{figure}

The use of our LDA-DFT scheme is illustrated by a study of the
edge reconstruction of the quantum
dot as a function of magnetic field strength. As is known from
Hartree-Fock and exact 
diagonalizations[\cite{MacD_Yang,Chamon,Yang_MacDonald_Johnson,Brey,Chklovskii2}], 
for strong confinement the 
quantum dot forms a maximum density droplet 
in which the density is uniform
at $\nu=1$ in the interior, and falls off rapidly to zero at 
$r\approx\sqrt{2N_{\rm el}}\ell_B=r_0$. 
As the magnetic field strength increases,
a ``lump'' of density breaks off, leaving a ``hole'' or deficit at about
$r=r_0$. This effect is due to the short-ranged attractive
exchange interaction:
it is energetically favorable to have a lump of density break off so that the
system can take advantage of the exchange energy in the lump. As $B$ is
further increased, the correlations will cause
incompressible strips with densities $\nu=p/q$ to appear
[\cite{Beenakker,Chklovskii,Gelfand,Ferconi}] on the edges, 
and incompressible droplets to form in the bulk
at densities $\nu=p/q$. 
Figure \ref{reconstr} depicts various stages
of edge 
reconstruction obtained by us as the magnetic field strength is increased.
The value of 
$B$
for which the exchange lump appears compares very well with the value
found by De Chamon and Wen[\cite{Chamon}] 
in Hartree-Fock and numerical diagonalizations.
At higher fields still, incompressible strips appear
at the edges, and incompressible droplets are formed in the bulk.

\begin{figure}[t]
\setbox3=\hbox{
\epsfxsize 13.0 cm
\epsfbox{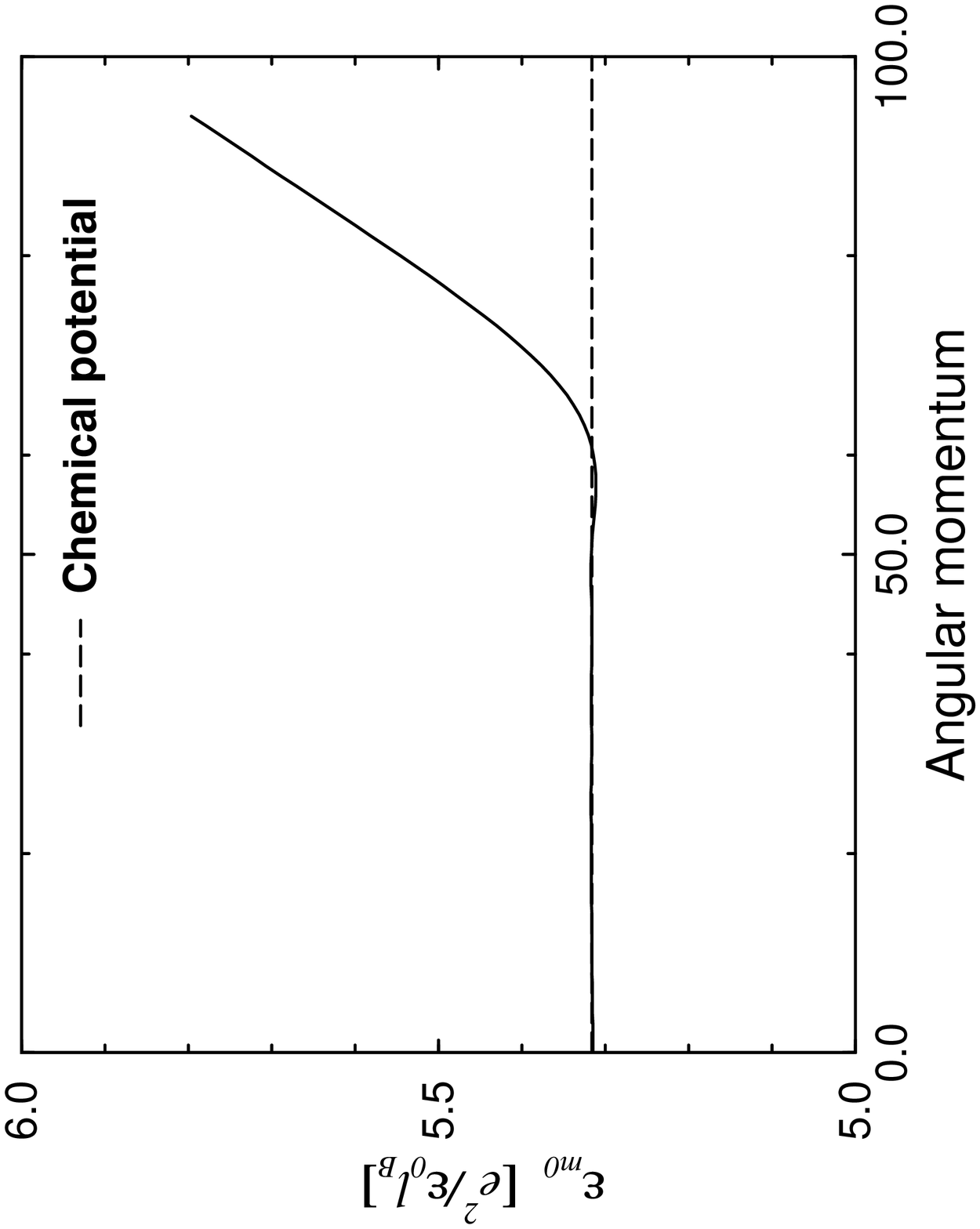}
}
\rotr3
\caption{Eigenvalues of the lowest-Landau level 
Kohn-Sham orbitals for $N_{\rm el}=40$ and
$B=4.10$ T as a function of angular momentum quantum number. The chemical
potential is indicated by the dashed line.}
\label{mu:fig} 
\end{figure}

Figure \ref{mu:fig} depicts the eigenvalues of the KS orbitals for
$N_{\rm el}=40$, and $B=4.10$ T. The dashed line indicates the chemical
potential of the system.
This figure then shows that all KS orbitals in the
bulk
are in fact degenerate. It may at first seem paradoxical that the eigenvalues
are degenerate on an incompressible strip, since, according to the picture
by Chklovskii, Shklovskii, and Glazman[\cite{Chklovskii}], 
on such a strip the density is constant, while
the total potential varies (since the electrons cannot screen the 
external potential). If the total potential varies, then ought not the
the eigenvalues of the KS orbitals localized on that strip vary, too, 
since these then in general are subjected to different potential energies?
The problem with this argument as applied to DFT is that it ignores the
effect of the exchange-correlation potential. As the external and
Hartree potentials vary across the strip, the exchange-correlation potential
varies across its discontinuity so as to completely screen out the
external and Hartree potentials. The discontinuity in $V_{\rm xc}$ does
{\em not} mean that this potential is fixed at the lower limit of its
discontinuity while the density is fixed at an incompressible strip. What it
does mean, is that $V_{\rm xc}$ is free to achieve any value across its
discontinuity so as to completely screen out the external and Hartree
potentials. In this way, it is perhaps better to think of incompressibility
as the limit of a finite compressibility approaching zero. A strip
can then remain incompressible with constant density so long as $V_{\rm xc}$
can screen the external and Hartree potentials, so the width of the
incompressible strip is given by the distance over which the external plus
Hartree potentials varies an amount given by the energy gap associated with
the density at that strip. 
Also, all bulk KS states are degenerate at the chemical potential. When a
single particle is added, the chemical potential simply increases a small
amount, and all KS orbitals are again degenerate at the chemical potential.

\begin{figure}[t]
\setbox4=\hbox{
\epsfxsize 13.0 cm
\epsfbox{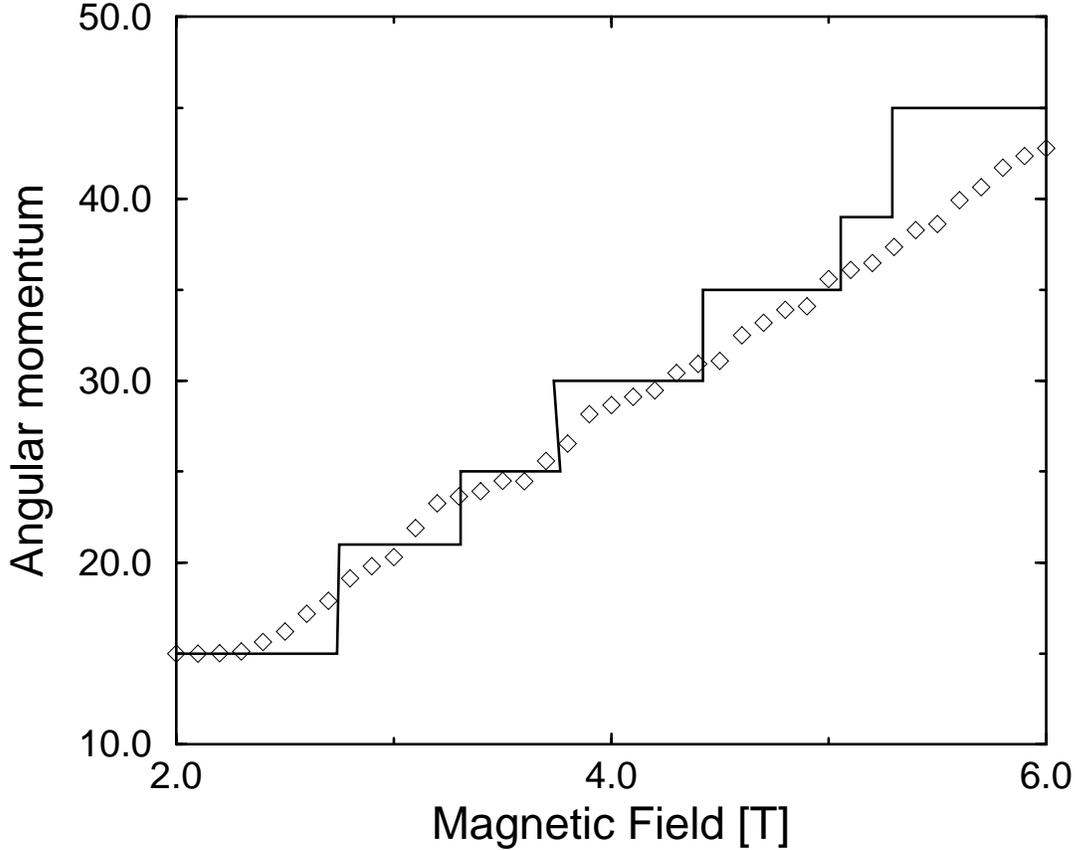}
}
\rotr4
\caption{Angular momentum {\em vs.} magnetic field strength for a six-electron
droplet in a parabolic potential with $\hbar\Omega=2.0$ meV. The solid line
is from numerical diagonalizations (Ref. 25), and the
diamonds are from our DFT-LDA calculations.}
\label{M_vs_B:fig}
\end{figure}

We have also tested the accuracy of our ensemble DFT-LDA approach by comparing
our results for a six-electron system in a confing parabolic potential
with the numerical diagonalization results by Yang, MacDonald and 
Johnson[\cite{Yang_MacDonald_Johnson}]. Figure \ref{M_vs_B:fig} depicts
angular momentum {\em vs.} magnetic field strength for this system. For better
comparison with the numerical diagonalizations, we used here only basis states
int the lowest Landau level ($n=0$). There
are clear plateau structures in the angular momenta, and readily
identifiable transitions, such as the initial instability of the
maximum density droplet at about 2.8 T, and a formation of a $\nu=1/3$ droplet at
about 5.3 T. The agreement
is in general very good, in particular considering that (a) the DFT
calculations are not constructed to give quantized angular momentum, (b) there
are no adjustable parameters in our approach, and (c) the numerical
diagonalizations used the full Coulomb interaction, while our DFT approach
only included the cusps at $\nu=1/3,2/5,3/5$, and $\nu=2/3$. The 
Levesque-Weiss-MacDonald interpolation formula used here tends to overestimate
the magnitude of
the exchange-correlation potential around $\nu=1/2$, which increases the values
at which transitions occur. For example, the $\nu=1/3$ droplet formation
occurs at 5.29 T in the numerical diagonalization, but at about 5.5T in
the DFT-LDA calculation. The overestimation of the exchange-correlation
potential also leads to decreased angular momenta at the different plateaus. For
example, while the $\nu=1/3$ droplet has angular momentum 45 in the
numerical diagonalization, examination of the density profiles shows that
the formation of a $\nu=1/3$ droplet occurs at an angular momentum
of about 40 in our calculations. We are presently
working on improving the exchange-correlation energy for better agreement
with numerical diagonalization. Initial calculations give highly
accurate results for the maximum density droplet instability, and the 
formation of a $\nu=1/3$ droplet.

\section{Including spin degrees of freedom}

As mentioned above, due to the small effective Land\'e $g$-factor in
GaAs, the spin degree of freedom play an important role in quantum
Hall systems, in spite of the large applied magnetic fields. We have 
begun to generalize our ensemble density functional approach to include
the spin degree of freedom. In general[\cite{vonBarth_Hedin}], 
spin density functional theory
has to be based on the single-particle density matrix
$\rho_{\sigma\sigma'}({\bf r})=\langle \hat\psi^\dagger_\sigma({\bf r})
\hat\psi_{\sigma'}({\bf r})\rangle$, where $\hat\psi_\sigma({\bf r})$
is the annihilation operator for an electron of spin $\sigma$ at
position ${\bf r}$. However, in the presence of a uniform external
magnetic field ${\bf B}=B\hat z$, the $z$-component of total electron
spin, $\hat S_z$ commutes with the Hamiltonian, and it is a reasonable
approximation to take $\rho_{\sigma\sigma'}({\bf r})$ to be
diagonal in the spin indices, $\rho_{\sigma\sigma'}({\bf r}) 
=n_\sigma({\bf r})\delta_{\sigma\sigma'}$, 
with $n_\sigma({\bf r})$ the up- and down-spin
densities. (Note, however, that this restriction
has to be lifted in order to study skyrmion-like spin textures, which
complicates the formalism a great deal.) We now obtain two sets of 
KS equations, one for each spin direction:
\begin{eqnarray}
&&\left\{\frac{1}{2m^*}\left[{\bf p}+\frac{e}{c}{\bf A}({\bf r})\right]^2
+V_{\rm ext}({\bf r})+V_{\rm H}({\bf r})+V_{{\rm xc},\sigma}({\bf r},{\bf B})
+\sigma g^*\mu_0 B\right\}\psi_{mn\sigma}({\bf r})\nonumber\\
&=&
\epsilon_{mn\sigma}\psi_{mn\sigma}({\bf r}).
\label{eq:spin_DFT}
\end{eqnarray}
Here, $\sigma=\pm1$, $\mu_0$ is the Bohr mangeton, and in the local
spin density approximation (LSDA) the exchange-correlation potentials
are
\begin{equation}
V_{{\rm xc},\sigma}({\bf r},{\bf B})=\left.
{\partial\over\partial n_\sigma}\left[n\epsilon_{\rm xc}(n_\uparrow,
n_\downarrow,B)\right]\right|_{n_\sigma=n_\sigma({\bf r})}.
\label{eq:VxcLSDA}
\end{equation}
It is more convenient to work with filling factor $\nu({\bf r})=2\pi\ell_B^2
n({\bf r})$ and spin polarization $\xi=(n_\uparrow-n_\downarrow)/
(n_\uparrow+n_\downarrow)$, in terms of which we have
\begin{eqnarray}
V_{{\rm xc},\uparrow} & = & {\partial\over\partial \nu}(\nu\epsilon_{\rm xc})
+(1-\xi){\partial\over\partial\xi}\epsilon_{\rm xc},\nonumber\\
V_{{\rm xc},\downarrow} & = & {\partial\over\partial \nu}(\nu\epsilon_{\rm xc})
-(1+\xi){\partial\over\partial\xi}\epsilon_{\rm xc},
\label{eq:Vxc_LSDA_2}
\end{eqnarray}
where now the exchange-correlation energy per particle in a homogeneous
system with a filling factor $\nu$ and polarization $\xi$ has to be
approximated. Except for a few data points obtained by small system
numerical diagonalizations[\cite{Chakraborty}], this is largely unkown.
In order to obtain a useful approximation, we start by considering
only the exchange energy $E_x[\nu_\uparrow,\nu_\downarrow]$. Since
the exchange interaction only couples electrons with parallel spins,
we have
\begin{equation}
E_x[\nu_\uparrow,\nu_\downarrow]=\frac{1}{2}E_x[\nu_\uparrow,\nu_\uparrow]
+\frac{1}{2}E_x[\nu_\downarrow,\nu_\downarrow].
\label{eq:exchange_1}
\end{equation}
In two dimensions, the exchange energy scales as $n^{3/2}$, so using
Eq. (\ref{eq:exchange_1}), we can correctly interpolate between
a fully polarized systems ($\xi=1$) and a completely un-polarized one
($\xi=0$) by writing
\begin{equation}
\epsilon_x[\nu,\xi]=\epsilon_x(\nu,\xi=1)+\left[
\epsilon_x(\nu,\xi=0)-\epsilon_x(\nu,\xi=1)\right]f(\xi),
\label{eq:exchange_2}
\end{equation}
where 
\begin{equation}
f(\xi)={(1+\xi)^{3/2}+(1-\xi)^{3/2}-2\sqrt{2}\over 2-2\sqrt{2}}.
\label{eq:f_xi}
\end{equation}
We then use the {\em same} interpolation for the correlation energy 
(excluding
the cusps for a moment), as
was first done by von Barth and Hedin[\cite{vonBarth_Hedin}] for
the three-dimensional electron gas in zero magnetic field, and write
\begin{eqnarray}
\epsilon_{\rm xc}'(\nu,\xi)&=&\epsilon_{\rm xc}^{\rm LWM}(\nu,\xi=1)
+\left[\epsilon_{\rm xc}(\nu,\xi=0)-\epsilon_{\rm xc}(\nu,\xi=1)
\right]f(\xi)\nonumber\\
&\equiv&\epsilon_{\rm xc}^{\rm LWM}(\nu,\xi=1)
+\delta\epsilon_{\rm xc}(\nu)f(\xi).
\label{eq:exc_spin_1}
\end{eqnarray}
The function $\delta\epsilon_{\rm xc}(\nu)$ can the be obtained
by calculating the energy difference between polarized and un-polarized
systems using data obtained from small system numerical diagonalizations
[\cite{Chakraborty}]. Accurate value for the exchange-correlation energy
for $\xi=0$ and $\nu=1$ is not available, but it is a useful approximation
to assume that the different spin directions are completely uncorrelated
at $\xi=0$ and $\nu=1$, which gives $\epsilon_{\rm xc}(\nu=1,\xi=0)
=\epsilon_{\rm xc}(\nu=1/2,\xi=0)$. This approximation also gives
a good value for the exchange enhancement, which is the energy required
to flip a spin in a polarized system at $\nu=1$.

So far, we have constructed a function $\epsilon_{\rm xc}'(\nu,\xi)$ which
gives a smooth interpolation for the exchange-correlation energy for
any value of $\nu$ and $\xi$. What is left is to add the cusps to this
function. We already have a good approximation for this at $\xi=1$.
We now need to extend this approximation to arbitrary values of $\xi$.
Very little is known about the cusps, {\em i.e.,} the energy gaps, for
arbitrary polarizations. It is known that there {\em is} a gap for 
un-polarized systems at fillings $\nu=2/5$, $\nu=3/5$, and $\nu=2/3$.
The gap, and thus the cusps, occur at very special `magic' configurations
at which the system can take advantage of a particularly low
correlation energy. Therefore, it seems plausible that for a given
value of $\nu$, say $\nu=2/5$, there cannot be an energy gap for any
value of $\xi$ between 0 and 1. In order to incorporate this assumption
into a usable approximation, we interpolate our cusp energy constructed
for polarized systems,
$\epsilon_{\rm xc}^{\rm C}(\nu)$, to arbitrary polarizations by 
multiplying it by a function $g(\xi)$ which is unity at $\xi=0$ and
$\xi=1$, and vanishes away from these values of polarization. All
together, then, we have
\begin{equation}
\epsilon_{\rm xc}(\nu,\xi)=\epsilon_{\rm xc}^{\rm LWM}(\nu)
+\delta\epsilon_{\rm xc}(\nu)f(\xi)+\epsilon_{\rm xc}^{\rm C}(\nu)g(\xi).
\label{eq:exc_spin_2}
\end{equation}
Figure \ref{V_xc} depicts $V_{{\rm xc}\sigma}$ (here for $\xi=1$)
used in our calculations.

We have applied this spin ensemble DFT to study the phase diagram of
a maximum density droplet. For large values of the Land\'e-factor $g$, the
maximum density droplet is fully polarized, and as the magnetic
field is increased, there is an instability to forming a spin-polarized
exchange-hole. But for small values of $g$, the instability is towards
forming a spin structure at the edge. The value of $g$ separating
the spin-polarized and spin-structured instabilities, 
$\widetilde g=g\mu_BB/(e^2/\epsilon_0\ell_B)\approx0.05$, is in good agreement
with the value found for
$\widetilde g\approx0.03$ found from numerical diagonalizations of parabolic dots
by
Yang, MacDonald and Johnson[\cite{Yang_MacDonald_Johnson}], and consistent with the
value obtained in
calculations by Kivelson {\em et al.}[\cite{Kivelson}], 
who used a Hartree-Fock
approximation in which the spin axis was allowed to tumble. They
obtained a value $\widetilde g=0.17$ for an infinite Hall bar.

\section{Conclusion and summary}
In conclusion, we have showed that ensemble density functional theory can
be applied to the FQHE. This opens the door to doing realistic calculations
for large systems. We believe that our results are also significant in that
they are the first LDA-DFT calculations of a strongly correlated system
in a strong magnetic field, and they are (to the best of our knowledge)
the first practical ensemble DFT calculations. We find excellent
agreement between our ensemble DFT calculations and numerical
diagonalizations and Hartree-Fock calculations. Preliminary
calculations including spin degree of freedoms are consistent
with numerical diagonalizations and Hartree-Fock calculations. 
We are presently working on improving the exchange-correlation
energy as a function of density and polarization. Finally, the
spin DFT presented here cannot be used to study skyrmions-like structures,
in which the spin density vector is tumbling in space. Work is
presently underway, together with J. Kinaret (Chalmers University of 
Technology) to develop such a theory. 

The authors would like to thank M. Ferconi, M. Geller and G. Vignale for helpful
discussions and for sharing their results prior to publications, 
K. Burke and E.K.U Gross for useful comments about the DFT, and M. Levy
for a discussion about ref. [\cite{Levy}]. O.H. would like to 
thank Chalmers Institute of Technology, where part of the numerical work was done.
This work was supported by
the NSF through grants DMR93-01433 and DMR96-32141.

\end{document}